\documentstyle[aps,prl,preprint,floats,epsfig]{revtex}

\newcommand{\dlnu}{D\ell\nu}
\newcommand{\dplnu}{D^+\ell^-\bar{\nu}}
\newcommand{\dzlnu}{D^0\ell^-\bar{\nu}}
\newcommand{\dslnu}{D^*\ell\nu}
\newcommand{\ddlnu}{D^{**}\ell\nu}

\newcommand{\bm}{B^-}

\newcommand{\bb}{B\bar{B}}

\newcommand{\cby}{\cos\theta_{B-D\ell}}

\begin{document}

\preprint{\tighten\vbox{\hbox{\hfil CLNS 98/1594}
                        \hbox{\hfil CLEO 98-18}
}}

\title{Measurement of the $B \rightarrow \dlnu$ Branching Fractions \\
and Form Factor}  

\author{CLEO Collaboration}
\date{\today}

\maketitle
\tighten

\begin{abstract} 
Using a sample of $3.3\times 10^6$ $B$-meson decays collected with
the CLEO detector at the Cornell Electron Storage Ring, we have studied $\bm\to\dzlnu$ and $\bar{B}^0\to\dplnu$ decays, where $\ell^-$ can be either $e^-$ or 
$\mu^-$. We distinguish $B \to \dlnu$ 
from other $B$ semileptonic decays by examining the net momentum and energy
of the particles recoiling against $D-\ell$ pairs. We find $\Gamma(B \to
\dlnu) = (14.1 \pm 1.0 \pm 1.2)\,{\rm ns}^{-1}$ and derive 
branching fractions for $B^- \rightarrow \dzlnu$ and 
$\bar{B^0} \rightarrow \dplnu$ of $(2.32 \pm 0.17 \pm 0.20)\%$ and $(2.20 \pm 0.16 \pm 0.19)\%$ respectively, where the 
uncertainties are statistical and systematic. We also investigate the $B \to
\dlnu$ form 
factor and the implication of the result for $|V_{cb}|$. 
\end{abstract}
\newpage

{
\renewcommand{\thefootnote}{\fnsymbol{footnote}}

\begin{center}
J.~Bartelt,$^{1}$ S.~E.~Csorna,$^{1}$ K.~W.~McLean,$^{1}$
S.~Marka,$^{1}$ Z.~Xu,$^{1}$
R.~Godang,$^{2}$ K.~Kinoshita,$^{2,}$%
\footnote{Permanent address: University of Cincinnati, Cincinnati, OH 45221.}
I.~C.~Lai,$^{2}$ P.~Pomianowski,$^{2}$ S.~Schrenk,$^{2}$
% ...
\end{center}
 
\small
\begin{center}
$^{1}${Vanderbilt University, Nashville, Tennessee 37235}\\
$^{2}${Virginia Polytechnic Institute and State University,
Blacksburg, Virginia 24061}\\
% ...
\end{center}

\setcounter{footnote}{0}
}
\newpage
The semileptonic decays of the $B$-meson play important roles in heavy quark 
physics. They provide our best information on the CKM matrix elements
$V_{cb}$ and $V_{ub}$~\cite{bb:CKM} and reveal the dynamics of heavy quark 
decay in their form factors. 
Heavy Quark Effective Theory (HQET)~\cite{bb:HQET} provides a 
framework for calculating the form factors for $b \to c\ell\nu$ decays 
and suggests a reliable method for extracting $|V_{cb}|$ by predicting that 
QCD effects are small at the kinematic point
where the final-state meson is at rest relative to the initial meson 
(zero recoil). Most studies of the form factors and $|V_{cb}|$~\cite{bb:dslnures} have used 
the $B \to \dslnu$ decay because   
its differential branching fraction near the zero-recoil region
is large and the QCD effects have been calculated to the highest 
accuracy~\cite{bb:neubert}. 
There have been two recent studies
of the mode $B \to \dlnu$~\cite{bb:dlnu}~\cite{bb:dlnualeph}.
 Here we present
a new, more precise study of $B \to \dlnu$
using a different analysis method. 

We select $B$ charm semileptonic decays including 
$B \to \dlnu$, $B \to \dslnu$, 
$B \to \ddlnu$, and $B \to D^{(*)}\pi\ell\nu$
decays
by identifying events with a $D$ ($D^0$ or $D^+$ and their charge conjugates)
 and a lepton.
We then separate
$B\to \dlnu$ from the other semileptonic modes 
 using the net energy and momentum of the
particle or particles recoiling against the $D-\ell$ pair.
Information on the partial width and differential decay rate
are obtained from the $B^- \to \dzlnu$ and $\bar{B}^0 \to \dplnu$
yields, which we extract in bins of the HQET variable
$w = (M_B^2 + M_D^2 - q^2)/(2M_BM_D)$, where $q^2$ is the squared invariant
mass of the virtual $W$.
 
The data used in this analysis were accumulated in the CLEO 
detector~\cite{bb:CLEO-nim}
at the Cornell Electron Storage Ring (CESR). The data consist of 
3.3 million $\bb$ events collected at the $\Upsilon(4S)$ resonance. We also use 1.6 ${\rm fb^{-1}}$ of data collected 60~MeV
below the $\Upsilon(4S)$ resonance to study the background from the 
$e^+e^- \to q\bar{q}$ continuum.
The CLEO detector measures the trajectories of charged particles in 
a drift chamber system inside a 1.5T superconducting solenoid.
The main drift chamber provides the specific ionization ($dE/dx$) of
charged particles and their time-of-flight (TOF) is provided
by scintillation counters surrounding the drift chamber. 
A CsI electromagnetic calorimeter is used in electron identification.
Muons register hits in counters embedded in steel surrounding the magnet.

In this analysis we select events having at least 5 charged tracks
and, to suppress non-$B\bar{B}$ events, the ratio of Fox-Wolfram 
moments~\cite{bb:fox}  $H_2/H_0 <$  0.45.
We reconstruct $D^0$ and $D^+$ candidates in the
decay modes $D^0 \to K^-\pi^+$ and $D^+ \to K^-\pi^+\pi^+$ respectively.
We distinguish $K$'s from $\pi$'s based on a $\chi^2$ probability
($P_K$ and $P_{\pi}$) that combines $dE/dx$ and TOF.  
Each daughter track must satisfy $P_i > 0.01$ and $P_i/(P_K + P_{\pi}) > 
0.3$, where $i$ is $\pi$ or $K$ as appropriate.
The candidate mass must be within 1.835~GeV $< m_{K\pi} <$ 1.893~GeV for 
$D^0$ and 1.846~GeV $< m_{K\pi\pi} <$ 1.890~GeV for $D^+$ decays. 
To suppress $D$-mesons produced in $e^+e^- \to c\bar{c}$ events, 
we require $|{\bf p}_D| < $  2.5~GeV/$c$. 

Electrons are identified using $dE/dx$, the shape
of the shower in the CsI calorimeter, and $E/p$, the ratio of the candidate's
energy deposit 
in the CsI to its momentum.
Electrons are required to have momenta between 0.8~GeV/$c$ and 2.4~GeV/$c$. 
Muons, selected within the same momentum window, 
must penetrate at least 5 interaction 
lengths of material.
This requirement places an implicit lower bound on the muon momentum
of about 
1.4~GeV/$c$. We require that the lepton have the same charge as the $K$ from 
the decay of the $D$-meson. 
Since the decaying $B$-meson is nearly at rest,
the $D$-meson and the lepton are in opposite hemispheres for more than 90\%
of $B \to \dlnu$ decays; we demand that this be so. 

For events satisfying these criteria, we compute $\cby$,
the cosine of the angle between the $D\ell$ momentum ${\bf p}_{D\ell} = {\bf p}_{D} + 
{\bf p}_{\ell}$ and the $B$
momentum ${\bf p}_B$, assuming that the decay is $B \to \dlnu$
and that the only missing particle is the massless neutrino; that is
\begin{equation}
\cos\theta_{B-D\ell} = \frac{2E_BE_{D\ell}-{M^2_B}-{M^2_{D\ell}}}{2|{\bf p}_B|
|{\bf p}_{D\ell}|}.
\end{equation}
This quantity lies 
between $-1.0$ and 1.0 for $B \to \dlnu$ decays. When final-state
particles are missing in addition to the $\nu$, as is the case for the
other $B$ semileptonic decay modes, $\cby$ decreases. We use
the distribution of $\cby$ to separate $B \to \dlnu$ events from
the other $B$ semileptonic decay modes after subtracting backgrounds.

The backgrounds
come from several sources: random $K\pi(\pi)$ combinations,
$D$-mesons matched
with a lepton from the other $B$ decay (uncorrelated), $D$-mesons 
combined with a lepton that is a granddaughter of the same $B$ (correlated), 
hadrons misidentified as leptons (fake lepton), 
and $e^+e^-\to q\bar{q}$ events. 

Events in the mass regions above and below
the $D$ peak (sideband) are utilized to estimate the random combination 
contribution.
A study using Monte Carlo-simulated $B$ decays shows that this method
subtracts the right amount of background within small uncertainties and that 
the $\cby$ distribution of the events in the sidebands reproduces that of the
background events in the signal region.

In uncorrelated background events, the angular distribution between the 
$D$ and $\ell$ is nearly uniform because they arise from different $B$-mesons,
both nearly at rest.  We take advantage of this uniformity: 
for each event in which the $D$ and $\ell$
are in the same hemisphere we reverse the lepton's direction and
compute $\cby$, thereby constructing this distribution for the
opposite-hemisphere background events.

The $e^+e^-\to q\bar{q}$ continuum background is measured using events 
collected off resonance.
The correlated background, which is small, arises from modes such as 
$B \to D_sD$ followed by $D_s \to X\ell\nu$ 
and $B \to DX\tau\nu$ followed by $\tau\to \ell\nu\bar{\nu}$.
We estimate these contributions using a  
Monte Carlo simulation. Fake lepton background is 
estimated by repeating the analysis using hadrons in place of 
leptons and then 
scaling the yield by the momentum-dependent electron and muon 
misidentification probabilities. 
Table~\ref{tab:events} summarizes the data yield and backgrounds.

\begin{table}[t]
\caption{Yields and Backgrounds.  Uncertainties are
statistical only.}
\label{tab:events}
\begin{tabular}{lcc}
                             & $D^0\ell$     & $D^+\ell$  \\    \hline
Total Yield                  & 12595 $\pm$ 112 & 18087 $\pm$ 134 \\ \hline
Random $K\pi(\pi)$ Combinations & 5083 $\pm$ 50 & 13502 $\pm$ 70 \\ 
Uncorrelated                 &  948 $\pm$ 63 &  761 $\pm$ 75 \\ 
Continuum                    &  452 $\pm$ 84 &  432 $\pm$ 104 \\ 
Correlated                   & 119 $\pm$ 16 &   119 $\pm$ 23 \\ 
Fake Lepton                  &   71 $\pm$ 19  &  26 $\pm$ 25  \\ \hline
Background-subtracted Yield  & 5922 $\pm$ 163 & 3247 $\pm$ 201 \\
\end{tabular}
\end{table}

\begin{figure}[t]
{\centerline{\hbox
{\epsfig{figure=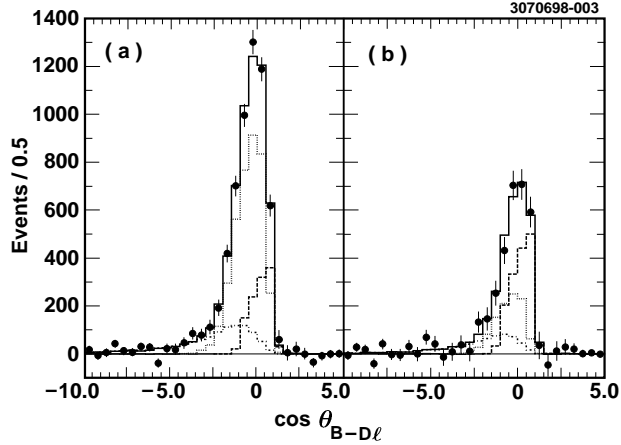,height=6.cm}}
}}
\caption{The $\cby$ distribution for (a) $B \to D^0X\ell\nu$  and 
(b) $B \to D^+X\ell\nu$. The data (solid circles) are 
overlaid with simulated 
$B \to \dlnu$ decays (dashed histogram),
$B \to \dslnu$ decays (dotted histogram),
$B \to \ddlnu + D^{(*)}\pi\ell\nu$ decays 
(dash-dotted histogram), and
their total (solid histogram). The normalizations of the simulated samples 
are provided by the fit.}
\label{fig:fit}
\end{figure}

\begin{figure}
{\centerline{\hbox
{\epsfig{figure=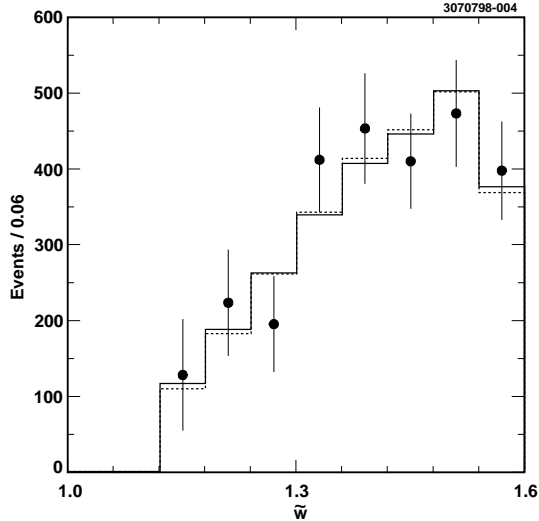,height=7.cm}}
}}
\caption{ The sum of $B^- \to \dzlnu$ and 
$\bar{B^0} \to \dplnu$ yields as a function of $\tilde{w}$,
for the data (solid circles) and using the best fit linear form factor 
(dashed histogram)
or dispersion relation inspired form factor of Boyd {\it et al.} 
(solid histogram).  }
\label{fig:fitff}
\end{figure}

After subtracting the backgrounds, we are left with $B \to D^0X\ell\nu$
and $B \to D^+X\ell\nu$ decays, where $X$ stands for zero or more pions or photons.
We divide each of these samples into ten equal bins of $\tilde w$ in the
range $1.0\leq \tilde{w}< 1.6$, where
$\tilde w$ is the reconstructed value of $w$, 
and is smeared by the detector resolution and motion of the
$B$.  In each  $\tilde w$ bin
we fit the $\cby$ distributions of the $B \to D^0X\ell\nu$
and $B \to D^+X\ell\nu$  samples 
for the yields of $\dlnu$, $\dslnu$ and the sum of
$\ddlnu$ and $D^{(*)}\pi\ell\nu$, using 
Monte Carlo-simulated $\cby$ 
distributions for each of these modes.
 In the simulation, we model $B \to \dlnu$ decays using ISGW2~\cite{bb:ISGW} 
and $B \to \dslnu$ decays using the form factors
measured by CLEO~\cite{bb:dslnuff}. We model the $D^{**}$   
mesons with radial and angular excitations using ISGW2
and non-resonant $D^{(*)}\pi$ states using the results of Goity and 
Roberts~\cite{bb:GR}. 
The detector simulation is based on 
GEANT~\cite{bb:geant}.
In our fits, we
apply the isospin symmetry constraint that the ratio of 
$B\to \dslnu$ to $B\to \dlnu$ decay rates should be the same 
for charged and neutral $B$-meson decays.  
Figure~\ref{fig:fit} shows the result
of a fit over all $\tilde{w}$. 
The sum of the $B^- \to \dzlnu$ and $\bar{B^0} \to \dplnu$
yields as a function of $\tilde{w}$ are displayed in 
Figure~\ref{fig:fitff}.
These $\tilde{w}$ distributions are the basis of our studies
of the $B^-\to \dzlnu$ and $\bar{B}^0\to \dplnu$ form factor and decay rate.

The differential
decay width of $B \to \dlnu$ is given by~\cite{bb:neubert}
\begin{equation}
\frac{d\Gamma}{dw} = \frac{G_F^2|V_{cb}|^2}{48\pi^3}(m_B+m_D)^2m_D^3(w^2-1)^{3/2}F_D(w)^2,
\end{equation}
where $G_F$ is the weak coupling constant, 
$m_D$ is the mass of the $D^0$ or $D^+$ and
$F_D(w)$ is the form factor.  
We fit the $B^-\to \dzlnu$ and $\bar{B}^0\to \dplnu$
yields in the intervals with $\tilde{w} > 1.12$ to extract
information on $|V_{cb}|$ and the form factor. 
The two bins with $\tilde{w}<1.12$ are excluded because they 
suffer from small rates due to the
 $(w^2-1)^{3/2}$ suppression and large backgrounds.
In our fit, the $\chi^2$ function for $B^- \to \dzlnu$ is expressed as
\begin{equation}
\chi^2_{\dzlnu} = \sum_{i=3}^{10}
\frac{[N_i^{obs} - \sum_{j=1}^{10}\epsilon_{ij}N_j]^2}
{\sigma_{N_i^{obs}}^2+\sum_{j=1}^{10}\sigma_{\epsilon_{ij}}^2N_j^2},
\end{equation}
where $N_i^{obs}$ is the yield in the $i$th $\tilde{w}$ bin and 
\begin{equation}
N_j = 2 f_{+-} N_{\Upsilon (4S)} {\cal B}_{K\pi}\tau_{B^-}\int_{w_j}dw \, 
d\Gamma/dw
\end{equation}
is the 
number of decays in the $j$th $w$ bin implied by the fit parameters.
Here $\tau_{B^-}$ is the $B^-$ lifetime~\cite{bb:blt}, 
${\cal B}_{K\pi}$ is the $D^0 \to K^-\pi^+$ branching 
fraction~\cite{bb:dbr}, $N_{\Upsilon (4S)}$ is the number of 
$\Upsilon (4S)$ events in the sample, and 
$f_{+-}$ is the $\Upsilon (4S)\to B^+ B^-$ branching fraction. 
An efficiency matrix, $\epsilon_{ij}$, accounts for reconstruction efficiency
and for the smearing of $\tilde{w}$.
The fraction of decays in each $w$ bin that are reconstructed
ranges between 17\% and 21\% and  the average 
$\tilde{w}$ resolution is 0.026, about one-half the bin width. The small
Monte Carlo statistical uncertainty in the efficiency matrix is represented by 
$\sigma_{\epsilon_{ij}}^2$.
We form $\chi^2_{\dplnu}$ analogously.
In the fits, we minimize $\chi^2 = \chi^2_{\dplnu} +
\chi^2_{\dzlnu}$, varying $|V_{cb}|F_D(1)$, the coefficients in
the parametrization of $F_D(w)/F_D(1)$, and $f_{+-}$.  We assume 
that the form factor parameters are common to $\dzlnu$ and $\dplnu$
decays and that
$B^+ B^-$ and $B^0 \bar{B}^0$ together saturate $\Upsilon (4S)$ decays. 

We investigate several parametrizations of the form factor.  Results
for all of these are summarized in Table~\ref{tab:ff}.  In all
fits we find $f_{+-}= 0.49 \pm 0.04$, consistent with previous
measurements~\cite{bb:fpm}, and similar correlation coefficients.
We first consider the common expansion $F_D(w)/F_D(1)= 
1 - \rho^2_D(w-1)+c_D(w-1)^2$.  When $c_D$ is constrained to be zero, we find
$\rho^2_D = 0.76 \pm 0.16$ and $|V_{cb}|F_D(1) = 0.0405 \pm 0.0045$ 
with the correlation coefficients
$\rho(|V_{cb}|F_D(1),\rho_D^2)=0.95$, $\rho(|V_{cb}|F_D(1),f_{+-})=0.12$ and
$\rho(f_{+-},\rho_D^2)=0.03$.  The $\chi^2$ is 8.8 for 13 degrees of
freedom.  When the $\dzlnu$ and 
$\dplnu$ samples are fit separately, they give consistent results for
all parameters.  
This form factor is superimposed on
the data in Figure~\ref{fig:fitff}.
When $c_D$ is allowed to vary, we find that it is consistent
with zero within large errors; that is, our data allow substantial
curvature but do not require it.  In this fit,
$\rho^2_D$ and $c_D$ are completely correlated
because our data are most precise at large values of $w$.

Dispersion relations constrain
the form factor.  Boyd $\it et$ $\it al.$\cite{bb:boyd} expand the 
form factor 
in the variable $z =  (\sqrt{w+1} - \sqrt{2N})/(\sqrt{w+1} + \sqrt{2N})$,
where $N \approx 1.1$.  
Because $z$ is small, this expansion converges more
rapidly than one in $w-1$.
Fitting for the linear coefficient $a_1$ with $N=1.108$, 
we find $a_1 = -0.043\pm 0.027$.
Expanding this form factor in powers of $w-1$ yields 
$\rho_D^2 = 1.30 \pm 0.27$ and $c_D = 1.21 \pm 0.31$ plus higher order terms.
Caprini {\it et al.}~\cite{bb:cap} have also used dispersion relations
to constrain the form factors. Their parametrization
leads to similar results.

\begin{table}[t]
\caption{Summary of the $B\to \dlnu$ form factor fits. In addition to the
quoted statistical uncertainties, there are 
fractional systematic uncertainties of 12\% in $\rho_D^2$ and 
$c_D$ and 8\% in $|V_{cb}|F_D(1)$.}
\label{tab:ff}
\begin{tabular}{lccccc} 
Form factor & $\rho^2_D$ & $c_D$ & $10^2|V_{cb}|F_D(1)$&$\chi^2/dof$ \\ \hline
Linear           & $0.76 \pm 0.16$ & $-$ & $4.05 \pm 0.45$ &8.8/13\\ 
Parabolic   &$0.77^{+1.18}_{-2.83}$&$0.01^{+1.70}_{-3.96}$&$4.05^{+1.51}_{-1.63}$&8.8/12\\
Boyd {\it et al.}$^*$$^\dag$&$1.30 \pm 0.27$ &$ 1.21\pm0.31$&$ 4.48\pm0.61$ & 8.9/13\\
Caprini {\it et al.}$^\dag$& $1.27\pm 0.25$ & $1.18\pm 0.26$ & $4.44\pm0.58$ & 8.9/13\\
\end{tabular}
$^*$ We find $a_1=-0.043\pm0.027$ for $N=1.108$. 

$^\dag$ This form factor also has terms of order $(w-1)^3$ and higher.
\end{table}

To obtain the $B \to \dlnu$ decay rate, we use the form factor
parameters provided by the fits and integrate $d\Gamma/dw$ over $w$.
The form factor of
Boyd {\it et al.}\cite{bb:boyd} gives
$\Gamma = (14.1 \pm 1.0)\,{\rm ns}^{-1}$.
The other parametrizations give the same result within 1\%.

The systematic uncertainties are given in Table~\ref{tab:syserr}. 
The uncertainties in the $B$-meson momentum 
and mass dominate because they respectively affect the width and mean 
of the $\cby$ distributions
and therefore the $\dzlnu$ and $\dplnu$ yields
extracted in each $\tilde{w}$ bin.
We have tuned our simulation to
reproduce the $B$ momentum distribution observed in fully reconstructed
$B$ decays; however, there is a 6~MeV uncertainty in the
mean and this leads to fractional 
systematic uncertainties of 
$5\%$ for $\rho^2_D$, $4\%$ for $|V_{cb}|F_D(1)$ and $3\%$ for $\Gamma$.  
The 1.8~MeV~\cite{bb:pdgmass} uncertainty in the $B$ mass 
generates uncertainties of 7\% for $\rho^2_D$, 4\% 
for $|V_{cb}|F_D(1)$ and 3\% for $\Gamma$.

The other large systematic error arises from uncertainty in the $\cby$ 
distribution of the combined $B \to \ddlnu$ and 
$B \to D^{(*)}\pi\ell\nu$ backgrounds. This distribution
depends mainly on the number of final-state pions that are not reconstructed.
We therefore separate it into two components: one in which the final-state 
$D$ is accompanied by one $\pi$ and the other in which it is accompanied
by two.
We vary their relative proportions from 1:4 to 2:3~\cite{bb:ALEPH} to evaluate
the systematic uncertainty.

The $B \to \dslnu$ form factors affect 
the distribution of these decays 
in $\cby$ and therefore influence the extracted $B \to 
\dlnu$ yield. We vary the
form factors within the uncertainties of the CLEO 
measurement~\cite{bb:dslnuff}, taking into account the correlations among the 
form factor parameters ($R_1, R_2$ and $\rho^2$).

The slope of the $B \to \dlnu$ form factor is varied in our fit 
so its uncertainty is included in the statistical error in the 
decay width. Using a linear form factor to extract the width rather
than the dispersion-relation-inspired form factor changes the width 
by 0.8\%.  The form factor can also affect the $\cby$ distribution used to
extract the $\dzlnu$ and $\dplnu$ yields in each $\tilde{w}$ bin
and the efficiency matrix.
Each of these effects is less than 1\%.

\begin{table}[t]
\caption{The fractional systematic uncertainties.}
\label{tab:syserr}
\begin{tabular}{lccc} 
Source  & $\rho^2_D$ & $|V_{cb}|F_D(1)$ & $\Gamma(B \to \dlnu)$ \\ \hline
Track-finding     & $-$ & 0.02& 0.035  \\ 
Lepton ID         & $-$ & 0.01& 0.020  \\
$K$ and $\pi$ ID  & 0.02 & 0.01 & 0.022 \\
Backgrounds       & 0.06 & 0.04 & 0.018 \\
$|{\bf p}_B|$ and $M_B$  & 0.08 & 0.05 & 0.042 \\
Luminosity        & $-$  & 0.01 & 0.018 \\
$\dlnu$ form factor & 0.01 & 0.01 & 0.010 \\
$\dslnu$ form factors & 0.01 & 0.01  & 0.005 \\
$\ddlnu$ model     & 0.04 & 0.03  & 0.026 \\ 
$D$ branching fractions & $-$  & 0.02 & 0.036 \\   
$\tau_B$          & $-$  & 0.02 & 0.026 \\ \hline
Total             & 0.11 & 0.08 & 0.085 \\ 
\end{tabular}
\end{table}

Our final result is
\begin{eqnarray}
\Gamma(B \to \dlnu) &=& (14.1 \pm 1.0 \pm 1.2)\,{\rm ns}^{-1}.
\end{eqnarray}
Multiplying this by the measured $B$-meson lifetimes gives the branching
fractions
\begin{eqnarray}
{\cal B}(B^- \to \dzlnu) &=&  (2.32 \pm 0.17 \pm 0.20)\% {\rm \,\,\, and}\\
{\cal B}(\bar{B^0} \to \dplnu) &=& (2.20 \pm 0.16 \pm 0.19)\%,
\end{eqnarray}
\normalsize
where the first errors are statistical and the second are systematic.
Since we derive both branching fractions
from the decay width, their errors are completely correlated.
This result is consistent with previous measurements
but is more precise. Combining it with the previous 
CLEO measurement~\cite{bb:dlnu},
taking into account statistical and systematic correlations, gives
\begin{eqnarray}
\Gamma(B \to \dlnu) &=& (13.4 \pm 0.8 \pm 1.2)\,{\rm ns}^{-1},\\
{\cal B}(B^- \to \dzlnu) &=&  (2.21 \pm 0.13 \pm 0.19)\%,
{\rm \,\,\,  and} \\
{\cal B}(\bar{B^0} \to \dplnu) &=& (2.09 \pm 0.13 \pm 0.18)\%,
\end{eqnarray}
where the errors in
the partial width and branching fractions are completely correlated.

Our studies of the form factor give
$\rho^2_D = 0.76 \pm 0.16 \pm 0.08$ (linear fit) and
$\rho^2_D = 1.30 \pm 0.27\pm 0.14$ and
$c_D = 1.21 \pm 0.31 \pm 0.15$ plus higher order terms  (dispersion relations).
The latter gives
\begin{equation}
|V_{cb}|F_D(1) = (4.48 \pm 0.61 \pm 0.37) \times 10^{-2}.
\end{equation}
Various authors have found $F_D(1) = 0.98 \pm 0.07$~\cite{bb:f1} and
$F_D(1) = 1.03 \pm 0.07$~\cite{bb:ISGW}, and a recent lattice calculation
finds the preliminary value
$F_D(1) = 1.069 \pm 0.029$~\cite{bb:lattice}. Using $F_D(1) = 1.0$,
we find $|V_{cb}| = 0.045 \pm 
0.006 \pm 0.004 \pm 0.005$, where the last uncertainty covers all of these 
values of $F_D(1)$. This value of $|V_{cb}|$ is consistent
with that from $B \to \dslnu$ decays, 
though its uncertainty is larger. 
Using the linear form factor, as has been done in most previous
studies of $B\to \dslnu$, gives a value of $|V_{cb}|$ that is
about 10\% smaller than this.  While the curvature
of the form factor is likely to have a smaller effect on
the $|V_{cb}|$ extracted from $B\to \dslnu$ decays,
its effect could nevertheless be important.

We thank C.Glenn Boyd and Matthias Neubert for useful discussions.
We gratefully acknowledge the effort of the CESR staff in providing us with
excellent luminosity and running conditions.
This work was supported by 
the National Science Foundation,
the U.S. Department of Energy,
Research Corporation,
the Natural Sciences and Engineering Research Council of Canada, 
the A.P. Sloan Foundation, 
the Swiss National Science Foundation, 
and the Alexander von Humboldt Stiftung.  

\nopagebreak

\end{document}